%Paper: hep-ph/9310301
%From: drtj@s-a.amtp.liv.ac.uk
%Date: Mon, 18 Oct 1993 10:43:21 +0100 (BST)

\input phyzzx
    \hsize=175mm
    \vsize=225mm
    \hoffset=-5mm
    \voffset=0mm
    \def\titleupset{\relax}
    \def\endtitlepage{\par
        \insert\footins{\floatingpenalty=20000 \vbox{\hbox{}\medskip
            \ialign{##\hfil\cr \the\Pubnum\crcr\the\date\crcr }
            \vskip -1.5\baselineskip }}
        \vfil \eject }

\newdimen\doublewidth
\newinsert\LeftPage
\count\LeftPage=0
\dimen\LeftPage=\maxdimen
\def \r{\rho}
\def \g{\tilde g}
\def \y{\tilde y}
\def\lfac{16\pi^2}
\def\pa{\partial}
\def \q{\quad}
\def \qq{\qquad}

\def\be{\beta}
\def\ga{\gamma}
\def\de{\delta}
\def\De{\Delta}
\def\ep{\epsilon}
\def\pole{{1\over{16\pi^2\ep}}}
\def\sig{\sigma}
\def\sigp{\sig '}
\def \la{\lambda}

\def\e{\ep^{\mu\nu\rho\sig}}
\def\psib{\overline{\psi}}

\def\ebar{\ep^{ijkl}}
\def\Dslash{D\!\!\!\! /}
%%%%%%%%%%%%%%%%%%%%%%%%%%%%%%%%%%%%%%%%%%%%%%%%
% macros for references
%
\def\us#1{\underline{#1}}
\def\ldf{\REF}
\def\nup#1({{\it Nucl.\ Phys.}\ $\us {B#1}$\ (}
\def\plt#1({{\it Phys.\ Lett.}\ $\us  {#1}$\ (}
%REFERENCES
\ldf\hvelt{G. 't~Hooft and M.~Veltman, \nup 44 (1972) 189.}
\ldf\dixon{J.A.~Dixon, preprint CTP-TAMU-69-92-REV.}
\ldf\siegel{W.~Siegel, \plt 84B (1979) 193.}
\ldf\Capper{D.M.~Capper, D.R.T.~Jones and P.~van Nieuwenhuizen,
\nup 167 (1980) 479.}
\ldf\siegelb{W.~Siegel, \plt 94B (1980) 37.}
\ldf\Leveille{D.R.T.~Jones and J.P.~Leveille, \nup 206 (1982) 473.}
\ldf\vlad{L.V.~Avdeev, G.A.~Chochia and A.A.~Vladimirov, \plt 105B (1981) 272.}
\ldf\Korner{ J.G.~K\"orner and M.M.~Tung, preprint MZ-TH/92-41.}
\ldf\Misiak{M.~Misiak, preprint TUM-T31-46/93.}
\ldf\Martin{S.P.~Martin and M.T.~Vaughn, preprint NUB-3072-93TH.}
\ldf\Tim{ D.R.T.~Jones (unpublished) 1979; see also
W.~Siegel, P.K.~Townsend and P~van Nieuwenhuizen, Proc. 1980 Cambridge meeting
on supergravity, ITP-SB-80-65.}
\ldf\hvand{ R.~van Damme and G. 't~Hooft, \plt 150B (1985) 133.}
\ldf\jack{I.~Jack, \plt 147B (1984) 405;
G.~Curci and G.~Paffuti, \plt 148B (1984) 78;
D.~Maison, \plt 150B (1985) 139.}
\ldf\jacko{I.~Jack and H.~Osborn, \nup 249 (1985) 472.}
\ldf\macka{M.E.~Machacek and M.T.~Vaughn, \nup 222 (1983) 83.}
\ldf\mackb{M.E.~Machacek and M.T.~Vaughn, \nup 236 (1984) 221.}
\ldf\mackc{M.E.~Machacek and M.T.~Vaughn, \nup 249 (1985) 70.}
%%%%%%%%%%%%%%%%%%%%%%%%%%%%%%%%%%%%%%%%%%%%%%%%%%%%%%%%%%%%%%%%%%%%%%%%
\Pubnum={LTH 320}
\date={}
\titlepage
\date={October 1993}
\title{\titleupset{Dimensional Reduction in Non-Supersymmetric Theories}}
\author{\titleupset{I. Jack, D.R.T. Jones and K.L. Roberts}}
\address{DAMTP, University of Liverpool, Liverpool L69 3BX }
%%%%%%%%%%%%%%%%%%%%%%%%%%%%%%%%%%%%%%%%%%%%%%%%%%%%%%%%%%%%%%%%%%%%%%%%%%%%%%
\abstract
It is shown that regularisation by dimensional reduction is a viable
alternative to dimensional regularisation in non-supersymmetric theories.
\endtitlepage

\chapter{Introduction}
Dimensional regularisation (DREG) is an elegant and convenient way of
dealing with the infinities that arise in quantum field theory beyond the tree
approximation.
\refmark{\hvelt}
It is well adapted to gauge theories because it preserves gauge invariance;
it is less well-suited, however, for supersymmetry
because invariance of an action
with respect to supersymmetric transformations only holds in general for
specific values of the space-time dimension $d$.
This is essentially due to the fact that a necessary
condition for supersymmetry is
equality of Bose and Fermi degrees of freedom.
In non-gauge theories it
is relatively easy to circumvent this problem,
and DREG as usually employed is, in fact, a supersymmetric procedure.
Gauge theories are a
different matter, however, and the question as to whether there
exists a completely satisfactory supersymmetric
regulator for gauge theories remains
controversial. This fact has been exploited recently to suggest that there
may be  supersymmetric anomalies.\refmark{\dixon}

An elegant attempt to modify DREG so as to render it
compatible with supersymmetry was
made by Siegel.\refmark{\siegel} The essential difference between Siegel's
method (DRED) and DREG is that the continuation from $4$ to $d$ dimensions
is made by {\it compactification\/} or {\it dimensional reduction}.  Thus
while the momentum (or space-time) integrals are $d$-dimensional in the
usual way, the number of field components remains unchanged and consequently
supersymmetry is undisturbed. (For a
pedagogical introduction to DRED see Ref.~\Capper).

As pointed out by Siegel himself, \refmark{\siegelb} there remain potential
ambiguities with DRED associated with treatment of the Levi-Civita symbol,
$\e$.
To see how these arise, recall that with DRED it would seem that necessarily
$d<4$,
 since the regulated action is, after all, defined by
dimensional {\it reduction}. Then, given $d<4$, one can define an
object $\ebar$ (where all the indices are now d-dimensional) and show
that algebraic inconsistencies result\refmark{\siegelb} unless $d=4$. A
related problem (of course) is the fact that the only consistent
treatment of $\ga^5$ within DREG is predicated\refmark{\hvelt}
 on having $d>4$; for a
discussion of this see Ref.~\Leveille , where it is suggested that one
may in fact continue the relations we need to describe anomalies from  $d>4$
to $d<4$ and so have our cake and eat it. There is
another (but again related) problem with DRED, arising from
the fact that in spite of the correct counting of degrees of freedom,
there are still ambiguities associated with
establishing invariance of the action. (Most apparent in component formalism:
see Ref.~\vlad ).
In spite of these reservations,
DRED remains the regulator of choice for supersymmetric theories, and
has survived practical tests to high loop levels.

In this paper, we address problems of a different nature,
which arise when DRED
is applied to non-supersymmetric theories.
That DRED is a viable alternative to DREG
was first claimed in Ref~\Capper.
Subsequently it has been adopted occasionally,
motivated usually by the fact that Dirac matrix algebra is easier
in four dimensions--and in particular by the desire to use Fierz identities.
For a recent example, see Refs.~\Korner --\Martin. As we shall see, however,
one must be very
careful in applying DRED to non-supersymmetric
theories because of the existence of
{\it evanescent couplings}. These were first described in Ref.~\Tim , and
independently discovered by van Damme and 't Hooft in 1984.\refmark{\hvand}
They argued, in fact,
that while DRED is a satisfactory procedure for supersymmetric
theories \refmark{\jack ,\jacko} (modulo the
subtleties alluded to above) it leads to a catastrophic loss of unitarity in
the non-supersymmetric case. Evidently there is an important
issue to be resolved here--is, as
the authors of Ref.~\hvand\ claim, use of DRED
forbidden (except in the supersymmetric case) in spite
of its apparent convenience? We shall claim that if DRED is employed in the
manner
envisaged by the authors of Ref.~\Capper, (which as we shall see differs in an
important
way from the definition of DRED primarily used in Ref.~\hvand ) then there is
no problem
with unitarity. We will present as evidence for this conclusion a set of
transformations
whereby the beta-functions of a particular theory (calculated using DRED) may
be related to the beta-functions of the same theory (calculated using DREG) by
means of coupling constant reparametrisation. The bad news is that a correct
description
of this (or any non-supersymmetric) theory impels us to a recognition of the
fundamental fact
that in general the evanescent couplings renormalise in a manner independent
and different
from the `` real '' couplings with which we may be tempted to associate them.
This means
that care must be taken as we go beyond one loop; nevertheless it is still
possible
to exploit the simplifications in the Dirac algebra which have motivated the
use of DRED.
We will return to this point later.

\chapter{Gauge theory with fermions}

We begin by considering a non-abelian gauge theory with fermions but no
elementary
scalars. The theory to be studied consists of a Yang-Mills multiplet $W^a_{\mu}
(x)$ with
a multiplet of spin $1\over 2$ Dirac\foot{the generalisation to two component
or
Majorana fields is straightforward} fields $\psi^{\alpha}(x)$ transforming
according to
an irreducible representation $R$ of the gauge group $G$.

The Lagrangian density ( in terms of bare fields ) is
$$
L_B = -{1\over 4}G^2_{\mu\nu} - {1\over{2\alpha}}(\pa^{\mu}W_{\mu})^2 +
C^{a*}\pa^{\mu}D_{\mu}^{ab}C^b +
i\psib^{\alpha}\ga^{\mu}D_{\mu}^{\alpha\be}\psi^{\be}
\eqn\AA
$$
where
$$
G^a_{\mu\nu} = \pa_{\mu}W_{\nu}^a - \pa_{\nu}W_{\mu}^a + gf^{abc}W_{\mu}^b
W_{\nu}^c
$$
and
$$
D_{\mu}^{\alpha\be} = \de^{\alpha\be}\pa_{\mu} - ig (R^a )^{\alpha\be}W_{\mu}^a
$$
and the usual covariant gauge fixing and ghost terms have been introduced.

The process of dimensional reduction consists of imposing that all field
variables depend only
on a subset of the total number of space-time dimensions- in this case $d$ out
of $4$ where
$d = 4 - \ep$. In order to fully appreciate the consequences of this procedure
we must then
make the decomposition
$$
W_{\mu}^a(x^j ) = \{ W_i^a (x^j ), W_{\sig}^a(x^j )\} \eqn\AB
$$
where
$$
\de^i_i = \de^j_j = d \qq \hbox{and}\qq \de_{\sig\sig} = \ep.
$$
It is then easy to show that
$$
L_B = L _B^d + L_B^{\ep} \eqn\AC
$$
where
$$
L _B^d = -{1\over 4}G^2_{ij} -{1\over{2\alpha}}(\pa^{i}W_{i})^2 +
C^{a*}\pa^{i}D_{i}^{ab}C^b + i\psib^{\alpha}\ga^{i}D_{i}^{\alpha\be}\psi^{\be}
\eqn\AD
$$
and
$$
 L_B^{\ep} = {1\over 2}(D_i^{ab}W^b_{\sig})^2 - g\psib\ga_{\sig}R^a\psi
W_{\sig}^a
-{1\over 4}g^2 f^{abc}f^{ade}W^b_{\sig}W^c_{\sigp}W^d_{\sig}W^e_{\sigp}.
\eqn\AE
$$
Conventional dimensional regularisation (DREG) amounts to using
Eq.~\AD\ and discarding
Eq.~\AE . The additional contributions from $ L_B^{\ep}$ are precisely what is
required to restore the supersymmetric Ward identities at one loop in
supersymmetric theories, as
verified in Ref.~\Capper .\foot{  Of course in simple applications it is in
general more
convenient to eschew the separation performed above and  calculate with
4-dimensional
and $d$-dimensional indices rather than $d$-dimensional and $\ep$-dimensional
ones.}

We would now like to rewrite Eq.~\AD\ and Eq.~\AE\ in terms of renormalised
quantities. This is usually done by simply rescaling all fields and coupling
constants. It is clear, however, from the dimensionally reduced form of the
gauge transformations:
$$\eqalign{
\de W^a_i &= \pa_i\Lambda^a + gf^{abc}W^b_i\Lambda^c\cr
\de W^a_{\sig}&= gf^{abc}W^b_{\sig}\Lambda^c\cr
\de\psi^{\alpha} &= ig(R^a )^{\alpha\be}\psi^{\be}\Lambda^a\cr}\eqn\AF
$$
that each term in Eq.~\AE\ is separately invariant under gauge
transformations. The $W_{\sig}$-fields behave exactly like scalar
fields, and are hence known as $\ep$-scalars.
The significance of this is that there is therefore no
reason to expect the $\psib\psi W_{\sig}$ vertex to renormalise in the same
way as the $\psib\psi W_i$ vertex (except in the case of supersymmetric
theories). In the
case of the quartic $\ep$-scalar coupling the situation is more complex since
in general of course more than one such coupling is permitted by Eq.~\AF . In
other words, we cannot in general expect the $f-f$~tensor structure
present in Eq.~\AE~ to be preserved under renormalisation. This is clear from
the abelian
case, where there is no such quartic interaction in $L_B^{\ep}$ but there is
a divergent contribution at one loop from a fermion loop.

At this point we have a choice. On the one hand, we could decide that it
doesn't matter if
Green's functions with external $\ep$-scalars are divergent( since they are
anyway unphysical)
and introduce a common wave function subtraction for $W_i$ and $W_{\sigma}$, a
wave function
subtraction for $\psi$ and a coupling constant subtraction for $g$, these all
being determined
(as usual) by the requirement that Green's functions with real particles be
rendered finite.
This is the procedure adopted in the main by  van~Damme and 't~Hooft. On the
other hand
we could insist on all Green's functions being finite, leading ineluctably to
the introduction
of a plethora of new subtractions or equivalently coupling constants. We argue
strongly that
it is only the latter procedure which has a chance of leading to a consistent
theory; the
former manifestly leads to a breakdown of unitarity (which in fact is the
conclusion
reached in Ref.~\hvand\ ). We proceed now to a
discussion of the renormalisation of Eq.\AC\ : conducted
in a somewhat old-fashioned way, in the interest
(we hope) of clarity.

We are therefore led to consider the following expressions for renormalised
quantities
$L^d$ and $L^{\ep}$ :
$$\eqalign{
L^d = &-{1\over 4}Z^{WW} (\pa_i W_j - \pa_j W_i )^2
-{1\over{2\alpha}}(\pa^{i}W_{i})^2\cr
      &-Z^{WWW}gf^{abc}\pa_i W_j^a W^{bi} W^{cj} -
                                  {1\over 4}Z^{4W}g^2f^{abc}f^{ade}W_i^b
W_j^cW^{di} W^{ej}\cr
      &+Z^{C^* C}\pa^i C^*\pa_i C + Z^{C^* CW}gf^{abc}\pa^i C^{a*}W_i^b C^c\cr
      &+Z^{\psib\psi}i\psib\ga^i\pa_i\psi + Z^{\psib\psi W}g\psib R^a\ga^i\psi
W^a_i\cr}\eqn\AFA
$$
 and
$$\eqalign{
L^{\ep}  = &{1\over 2}Z^{\ep\ep}(\pa_i W_{\sig})^2
            + Z^{\ep\ep W} gf^{abc}\pa_i W_{\sig}^a W^{bi} W_{\sig}^c\cr
           &+Z^{\ep\ep WW}g^2f^{abc}f^{ade}W_i^b W_{\sig}^c W^{di} W_{\sig}^e
           -Z^{\psib\psi\ep}h\psib R^a\ga^{\sig}\psi W^a_{\sig}\cr
           &-{1\over 4} \sum_{r=1}^p Z_r^{4\ep}\la_r
            H^{abcd}_rW^a_{\sig}W^c_{\sigp}W^b_{\sig}W^d_{\sigp}.\cr}\eqn\AFB
$$
Eq.~\AFA\ is the usual expression for the Lagrangian in terms of
renormalised parameters. The labelling of the various subtraction constants is
not particularly conventional but (we hope) self-explanatory. In Eq.~\AFB\ we
have introduced a
``Yukawa'' coupling $h$ and a set of $p$ quartic couplings $\la_r$. The number
$p$ is given
by the number of independent rank four tensors $H^{abcd}$ which
are non-vanishing when symmetrised with respect to $(ab)$ and $(cd)$
interchange. In $SU(2)$,
this number is $2$: $\de^{ab}\de^{cd}$, and
$\de^{ac}\de^{bd} + \de^{ad}\de^{bc}$.  In $SU(3), p=3$, and for $SU(N)$
($N>3$),
$p=4$. The $\la_r$ mix non-trivially under renormalisation; it is
straightforward (but tedious)
to calculate
their one-loop $\be$-functions in, for example, the $SU(N)$ case and check that
for a supersymmetric
theory the $f-f$ tensor structure is preserved. We will not tax the reader's
patience by presenting
these results; but in the next section all the $\be$-functions are calculated
for a particular
$SU(2)$ theory.

The results for some of the subtraction constants at one loop are as follows:

$$\eqalign{
Z^{WW} &= 1 + \pole g^2[({13\over 3} - \alpha )C_2 (G) - {8\over 3}T(R)] \cr
Z^{WWW} &= 1 + \pole g^2[({17\over 6} - {3\over 2}\alpha)C_2 (G) - {8\over
3}T(R)] \cr
Z^{\psib\psi} &= 1 - \pole g^2[2\alpha C_2  (R)] \cr
Z^{\psib\psi W} &= 1 - \pole g^2[ {{3+\alpha}\over 2}C_2 (G) + 2\alpha C_2 (R)]
\cr}\eqn\AG
$$
and
$$\eqalign{
Z^{\ep\ep} &= 1  + \pole [g^2 (6-2\alpha )C_2 (G) - 4h^2 T(R)]\cr
 Z^{\ep\ep W} &= 1 + \pole [g^2{{9-5\alpha}\over 2}C_2 (G) - 4h^2 T(R)]\cr
Z^{\psib\psi\ep}&= 1 - \pole g^2 [ (6+2\alpha )C_2 (R) - (3 - \alpha )C_2 (G)]
-\pole h^2 [ 2C_2 (G) - 4C_2 (R)]\cr}\eqn\AH
$$
where
$$
\eqalign{
\de^{ab}C_2 (G) &= f^{acd}f^{bcd}\cr
\de^{ab}T(R) &= \hbox{Trace}[R^a R^b ] \qq\hbox{and}\cr
C_2 (R) 1 &= R^a R^a . }
$$
 At one loop, subtraction constants with no external $\ep$-scalars depend only
on
the real couplings while subtraction constants with external $\ep$-scalars
depend
on both real and evanescent couplings. Notice that we are using minimal
subtraction, so that all the Z's contain only poles in $\ep$.
{}From Eqs.~\AG~ and~\AH~ it is easy to verify
Slavnov-Taylor identities such as :
$$
{{Z^{WWW}}\over{Z^{WW}}} = {{Z^{\ep\ep W}}\over{Z^{\ep\ep}}}
= Z^g \sqrt{Z^{WW}}\eqn\HI
$$
where $Z^g$ is the renormalisation constant for $g$.

It is straightforward to calculate the one loop $\be$-function for $h$; the
result is :
$$
\be_h (g, h) = {h\over {\lfac}}[(4h^2 - 6 g^2 )C_2 (R) + 2h^2 T(R) - 2h^2 C_2
(G)]
\eqn\HJ
$$
which is to be contrasted with the result for $\be_g$, which is just as for
DREG:
$$
\be_g (g) = {g^3 \over{\lfac}}[ {4\over 3}T(R) - {11\over 3}C_2 (G)].
\eqn\HK
$$
The fact that $\be_g$ is independent of $h$ at one loop is a trivial
consequence
of minimal subtraction. It is interesting that this remains true at two
loops\foot{this
is clear from the calculations presented in Ref.~\Capper , although not
emphasised because
the distinction between $g$ and $h$ was not made in that paper.}: moreover the
result for
$\be_g^{(2)}$ is precisely the same as that obtained using DREG. We do not
know whether this persists to all orders. In section (3), however, we will find
that in a
more general theory (specifically one involving genuine scalar particles)
there {\it are\/} real couplings whose $\be$-functions depend on evanescent
couplings
beyond one loop.

We see that even if we choose to set $h = g$, the two $\be$-functions
are not identical, unless we also have that
$$
5C_2 (G) + 2T(R) = 6 C_2 (R). \eqn\HL
$$
The solution $ C_2 (G) = C_2 (R) = 2T(R) $ corresponds to a supersymmetric
theory. (The
awkward factor of 2 in front of $T(R)$ is due to our adoption of Dirac
fermions).
In general it is clear that the choice $h=g$ is not renormalisation group
invariant: that is, if it is made at one renormalisation scale it is not
true at another. Hence to investigate the structure of the
renormalised theory it is important that this choice (which has been
implicitly made in most DRED calculations) be {\it not\/} made .  We will
return later to
the issue of the validity of calculations involving $h = g$ and corresponding
choices for the
other evanescent couplings.

It is well known that different regulation procedures lead in general to
different results for
(amongst other things) $\be$-functions. We may expect, however, that the
$\be$-functions
obtained with two different regulators can be transformed into each other by
means of
coupling constant redefinition. It was this procedure which, for example,
established the
equivalence of the DRED and DREG results for the $N=4$
supersymmetric gauge theory. In Ref.~\hvand\
it is asserted that it is {\it not\/} possible to transform the DRED
$\be$-functions into
the DREG ones; both in general and in the context of a
particular model (which we call the DH model). In the next
section, we show how in fact there does exist such a transformation, as long as
we both
implement DRED in the manner described above and maintain the distinction
between real
and evanescent couplings. On the basis of this result we intend to argue that
DRED is
quite valid and (contrary to Ref~\hvand ) a perfectly valid alternative to DREG
in the
non-supersymmetric case.

\chapter{ The DH Model}
In this Section we show explicitly how the $\be$-functions evaluated using
our version of
DRED may be transformed into the $\be$-functions evaluated with DREG, with
reference to a specific example--namely the toy model
introduced in Ref.~\hvand.
As mentioned earlier, the version of DRED used in Ref.~\hvand\ is crucially
different from the one which we advocate, and in fact leads to $\be$-functions
which are not equivalent up to coupling constant redefinition to those of
DREG. We shall, however, maintain that our implementation of DRED is natural
and appropriate.

The toy model considered by van Damme and 't Hooft in Ref.~\hvand\ has gauge
group
$SU(2)$, a multiplet of Dirac fields $\psi$ and a multiplet of scalar fields
$\phi$ each of
which transforms according to the adjoint representation.The bare
Lagrangian is
$$
L=L_B^d+L_B^{\epsilon}\eqn\BA
$$
where
$$\eqalign{
L_B^d&=-{1\over
4}(G_{ij}^a)^2+{1\over2}(D_{i}\phi^a)^2+i{\bar\psi}^a\Dslash\psi^a
\cr
&\quad
+iy\epsilon^{abc}\bar\psi^a\phi^b\psi^c-{1\over8}\la(\phi^2)^2\cr}\eqn\BB
$$
and
$$\eqalign{
L_B^{\epsilon}&=+{1\over2}(D_iW_{\sigma}^a)^2-{1\over2}\rho_1(W_{\sigma}^a)^2
\phi^2
+{1\over2}\rho_2(W_{\sigma}^a\phi^a)^2 \cr
&\quad +ih\epsilon^{abc}\bar\psi^aW_{\sigma}^b\gamma_{\sigma}\psi^c -{1\over4}
\rho_4(W_{\sigma}^2)^2
+{1\over4}\rho_5(W_{\sigma}^aW_{\tau}^a)^2. \cr}\eqn\BC
$$
(We have omitted ghost and gauge fixing terms). The corresponding renormalised
Lagrangian $L=L^d+L^{\ep}$ is obtained in an
analogous fashion to that for the gauge theory with fermions in Section (2).
In particular, we have
$$
y \rightarrow Z^y y,\q \la\rightarrow Z^{\la}\la, \q h\rightarrow Z^h h,
\q \rho_i\rightarrow Z^{\rho_i}\rho_i,\qq i=1,2,4,5. \eqn\BD
$$
The one-loop $\be$-functions for the various couplings, both real and
evanescent, can be calculated by standard methods; the results are:
$$\eqalign{
\be^{(1)}_{\la} &= 11\la^2 - 24{\la}{\tilde g} +24{\tilde g}^2 +16{\la}y
-32{\tilde y}^2 \cr
\be^{(1)}_{\tilde y} &= 16{\tilde y}^2 - 24 {\tilde y} {\tilde g} \cr
\be^{(1)}_{\g} &= -{26\over3}{{\tilde g}^2} \cr
\be^{(1)}_{\rho_1}&= 8\rho_1^2 + 2\rho_2^2 + 4\rho_1\rho_4 - 8\rho_1\rho_5 +
2\rho_2\rho_5 \cr
&\q - 16{\rho_3}{{\tilde y}} + 4{\tilde g}^2
+ \rho_1(-24{\tilde g} + 8{\tilde y} + 8\rho_3) + 5\la\rho_1- \la\rho_2 \cr
\be^{(1)}_{\rho_2} &= -10\rho_2^2 + 16\rho_1\rho_2 + 4\rho_2\rho_4 -
2\rho_2\rho_5 -6{\tilde g}^2 \cr
&\q+\rho_2(-24{\tilde g} + 8{\tilde y}+ 8\rho_3) + 2\la\rho_2 \cr
\be^{(1)}_{\rho_3} &= \rho_3(-12{\tilde g}+16\rho_3) \cr
\be^{(1)}_{\rho_4} &= 16\rho_4^2 + 6\rho_5^2 - 16\rho_4\rho_5 - 16\rho_3^2 +
6{\tilde g}^2 \cr
&\q +\rho_4(-24{\tilde g}+16\rho_3)+ 6\rho_1^2 - 4\rho_1\rho_2 \cr
\be^{(1)}_{\rho_5} &= -14\rho_5^2 + 24\rho_4\rho_5 -
6{\tilde g}^2 + \rho_4(-24{\tilde g} + 16\rho_3) -
2 \rho_2^2 \cr}\eqn\BE
$$
where we have defined $\tilde y = y^2$, $\tilde g = g^2$ and $\rho_3 = h^2$.
Here and subsequently we suppress a factor
of ${(16\pi^2)}^{-L}$ in  the expression
for an $L$-loop  $\be$-function contribution.
Note that, as in section (2), the one-loop $\be$-functions for real couplings
do not depend on
the evanescent couplings due to the extra factor of $\ep$ associated with
the $\ep$-scalars, a fact which will simplify the consideration of
coupling constant redefinitions later.

We now want to compare the DRED and DREG results for the two-loop beta
functions.
At the two-loop level, we noted in section (2) that, for the class of theories
considered there, the DRED result for $\beta_g$ was independent of the
evanescent couplings;
 this property persists for the DH model and is clearly true in general. Let us
now consider the DRED calculation of $\be_\la^{(2)}$ and $\be_{\y}^{(2)}$.
There are two classes of two-loop diagrams contributing to the
renormalisation of real couplings; those which involve $\ep$-scalars
and those which do not. The set of diagrams which do not involve $\ep$-scalars
of course yield the same result as for dimensional regularisation. In
other words the difference between DREG and DRED arises solely from the
graphs with $\ep$-scalars, and consequently we shall limit our attention
to these. The calculation of the contributions to the
$\be$-functions from this class of graphs is rather straightforward, since the
presence of a factor of $\ep$ from the multiplicity of the $\ep$-scalars
means that we only need the double pole from the Feynman integral. Each graph
also requires corresponding counter-term diagrams, and
it is perhaps appropriate at this
point to explain the difference between our prescription for DRED, and that
adopted in Ref.~\hvand, since the difference resides in our treatment of the
counter-terms. According to our prescription, we construct the
counter-term diagrams for a graph with an
$\ep$-scalar loop on its own merits, by replacing
divergent sub-diagrams of the original diagram by counter-term insertions with
the same pole structure. This is equivalent to constructing one-loop
counter-term diagrams with insertions of $Z^{\la(1)}$, $Z^{\g(1)}$,
$Z^{\y(1)}$,
$Z^{\rho_i(1)}$.
 van Damme and 't Hooft , on the other hand, construct one-loop counter-term
diagrams by replacing counter-terms for evanescent couplings by those for the
corresponding real couplings, i.e. replacing $Z^{\rho_i(1)}$ by $Z^{\g(1)}$.
This prescription certainly seems incompatible with our
philosophy of taking the evanescent couplings seriously and distinguishing
them from real couplings, since it would not eradicate divergences from
graphs with external $\ep$-scalars.
With our prescription, however, some care needs to be exercised in determining
the correct counter-term diagrams,
since a sub-loop with $\ep$-scalars,
external real fields and a divergent Feynman
integral is nevertheless finite owing to the multiplicity-factor of $\ep$,
and hence does not require a subtraction.

The set of graphs to be calculated
may be easily obtained from Refs.~\macka - \mackc , by replacing one or more
vector propagators by $\ep$-scalar ones. The fact that
there is no $W_{\sig}-\phi - \phi$ vertex is a considerable simplification.
The results for the {\it difference\/} between DRED and DREG calculations
(i.e. $\de\be = \be_{DRED} - \be_{DREG}$) are as follows:
$$\eqalign{
\delta\be_{\g}^{(2)} = &0\cr
\delta\be_{\y}^{(2)} = &16\g^2\y \cr
\delta\be_\la^{(2)} = &-4\la(36\r_1^2 + 16\r_2^2 - 24\r_1\r_2) + 8(24\r_1^3 -
12\r_2^3 \cr
&+ 32\r_1\r_2^2 - 24\r_1^2\r_2) + 32\r_4(3\r_1^2 + \r_2^2 -
2\r_1\r_2)\cr  &- 32\r_5(6\r_1^2 + \r_2^2 -
4\r_1\r_2) - 96(3\r_1^2 + \r_2^2 - 2\r_1\r_2)\g \cr &+ 96(2\r_1-\r_2)\g^2
-16\g^3
-128(3\r_1-\r_2)\r_3\y \cr &+ 16\r_3(12\r_1^2 - 8\r_1\r_2 + 4\r_2^2) +
16\la\g^2 . \cr}\eqn\BF
$$
The fact that $\delta\be_{\y}^{(2)}$ is also independent of the evanescent
couplings is quite
remarkable, resulting from a cancellation of a large number of individual
contributions.
By itself, this result would have lent support to the  conjecture\refmark{\Tim}
that the real
coupling DRED $\be$-functions are indeed independent of the evanescent
couplings. The
conjecture is, however, laid to rest by $\delta\be_\la^{(2)}$. We now proceed
to show that the
DRED results are nevertheless equivalent to the DREG ones in the sense that
they may be
transformed into them by a finite perturbative reparametrisation of the
coupling constants.

Given a theory with coupling constants $\{\la_i\}$, if we define new couplings
$\{{\la_i}'\}$
by ${\la_i}' = \la_i + \de\la_i$ then the resulting change in the two loop
$\be$-functions
is given by
$$
\de\be_i = {\be_i^{(2)}}'(\la ) - \be_i^{(2)} (\la ) =
\be_j^{(1)} {\pa\over{\pa\la_j}}\de\la_i -
\de\la_j{\pa\over{\pa\la_j}}\be_i^{(1)}.\eqn\BG
$$
Our task is to demonstrate a set of $\de\la_i$ for the DH model such that the
resulting $\de\be_i$
precisely cancel Eq.~\BF\ , restoring the DREG results for the real $\beta$
functions.
This is a straightforward calculation (the tedium of which was ameliorated by
employment of
REDUCE) and the result is that the following set of $\de\la_i$ does the
business:
$$\eqalign{
\de\la &= {1\over 26}\Delta ( 11\la^2 + 24\g^2 -32\y^2 -24\la\g + 16\la\y )
+ 12\r_1^2 + 4\r_2^2 -8\r_1 \r_2\cr
\de\y &= {1\over 26}\Delta (16\y^2 - 24\g\y )\cr
\de\g &= {1\over 3}( 2 - \Delta )\g^2\cr}\eqn\BH
$$
where $\Delta$ is an arbitrary constant. It should be
emphasised that this is a non-trivial
result in that the existence of the solution {\it does\/}
depend on the precise values of the coefficients in Eq.~\BF.

It is clear that the required transformations become particularly simple if we
take $\De=0$ in Eq.~\BH. The resulting $\de\la$ and $\de\g$ then correspond
precisely to the potential finite contributions to $Z^{\la}\la$ and
$Z^{\g}\g$ which arise at one loop when a divergence from a Feynman integral
is multiplied by a multiplicity factor of $\ep$. (Of course, since we are
using minimal subtraction prescription we discard these finite contributions.)
This is intriguing, since a prescription mentioned {\it en passant\/} in
Ref.~\hvand\ (their ``system 3'') also led (for a general theory) to
$\be$-functions related to
those for DRED by the same transformation. This prescription is not
manifestly identical to ours, since the multiplicity factor for the
$\ep$-scalars is set to $\ep$ only at the end of the calculation--in other
words, from our point of view the counter-terms are not evaluated using minimal
subtraction; nevertheless we see from the above remarks that it leads to
identical results, at least for the DH model. It would be interesting to prove
the equivalence of these two prescriptions in general.

\chapter{Discussion}

The dependence on the evanescent couplings of the
DRED $\be$-functions seems to pose a serious problem. Most DRED users
have in fact dealt with the evanescent couplings by
setting them equal to the real couplings suggested by the bare Lagrangian;
in the DH model, for example, this corresponds to $\r_i = \g$ for all $i$. But
the evanescent couplings evolve differently, so  it would seem that
two ``observers'' testing physics at different energy scales could not both
make this
choice. We saw in the last section, however, that (at least for the DH model
through
two loops) the DRED $\be$- functions for the real couplings are in fact
equivalent to
to the DREG ones. Of course if DRED and DREG are really to describe the same
physics, it is important that the coupling constant redefinition that achieves
this also
renders the DRED S-matrix (for real particles) identical to that for DREG - in
particular,
independent of the evanescent couplings.  A simple example of this effect in
the DH model
is as follows. Consider the one loop contributions to the $\phi^4$ interaction
from
a pair of $\r_1$ and/or $\r_2$ vertices. These generate finite contributions to
the
vertex and so it is quite clear that with DRED the cross-section for
$\phi-\phi$ scattering,
for example, depends on $\r_1$ and $\r_2$. But it is easy to check that the
redefinition of
$\la$ from Eq.~\BH\ is precisely what is required to remove these
contributions. So when we
start with DRED and implement  Eq.~\BH\ not only are the resulting
$\be$-functions the
same as the DREG ones, but the resulting S-matrix is also identical. The
evanescent sector
is completely decoupled, and goes away as $\ep\rightarrow 0$, because we
have been careful to make Green's functions with external scalars finite.

Now DRED is only really useful in non-supersymmetric
theories if we can calculate without
splitting off the $\ep$-scalars, so that
Dirac algebra can be carried out in four
dimensions. This involves setting
the evanescent couplings equal to their ``natural''
values, as we described above. It should now be clear, however, that this
procedure is quite
harmless, as long as sub-divergences are subtracted at the level of the
integrals,
instead of by means of counter-term insertions.
 Great care would be needed however, if comparison between calculations
performed with different values of the renormalisation scale $\mu$ were
to be desired.

In conclusion, we have argued that DRED as it has been customarily employed
is perfectly valid. We have demonstrated
this through two loops in a simple model;
 but it seems clear that the conclusions are general: since both DRED and DREG
start from the same bare theory, it is inevitable that the
corresponding renormalised theories are equivalent physically, as
indeed we find. Nevertheless, we feel that it is important that users
of DRED be aware of the existence and
significance of the evanescent couplings.
\medskip
\noindent\undertext{Acknowledgements}: This work
was begun during a visit by one
of us (TJ) to the Aspen Center for Physics. Thanks go to
Howie Haber for drawing  attention
to Ref.~\Misiak, and to Martin Einhorn for comments.  IJ and KLR were supported
by the SERC via an
Advanced Fellowship and a Research Studentship respectively.
\refout
\bye